\pdfoutput=1
\documentclass{llncs}
\usepackage{subfigure}
\usepackage{graphicx}
\usepackage{amssymb}
\usepackage[algoruled,linesnumbered]{algorithm2e}

 \let\itPi\Pi \let\itGamma\Gamma \let\itTheta\Theta \let\itOmega\Omega
 \def\Pi{\mathrm{\itPi}}
 \def\Gamma{\mathrm{\itGamma}}
 \def\Theta{\mathrm{\itTheta}}
 \def\Omega{\mathrm{\itOmega}}

\catcode`\"=\active
\def "{\begingroup``\def "{\endgroup''}}

\spnewtheorem{probl}{Problem}{\bfseries}{}
\def\ex#1{{\it Example \##1: }}

\def\m{^{\raise-1pt\hbox{-}}}

\let\ro\overrightarrow
\let\lo\overleftarrow

\def\G{{\cal G}}
\def\H{{\cal H}}
\def\cand{\mathop{\hbox{cand}}}
\def\prog#1{{\tt #1}}

\begin{document}

\title{A New Approach to the Small Phylogeny Problem\\ (Technical Report)}
\author{Jakub Kov\'a\v{c}\inst{1}
        \and Bro\v{n}a Brejov\'a\inst{1}
        \and Tom\'a\v{s} Vina\v{r}\inst{2}}
\institute{Department of Computer Science,
           Faculty of Mathematics, Physics, and Informatics,
           Comenius University, Mlynsk\'a Dolina,\\842~48 Bratislava, Slovakia,
           e-mail: {\tt kuko@ksp.sk}, {\tt brejova@dcs.fmph.uniba.sk}
           \and
           Department of Applied Informatics,
           Faculty of Mathematics, Physics, and Informatics,
           Comenius University, Mlynsk\'a Dolina,\\842~48 Bratislava, Slovakia,
           e-mail: {\tt vinar@fmph.uniba.sk}}
\maketitle

\begin{abstract}
In the small phylogeny problem we, are given a phylogenetic tree and gene orders of
the extant species and our goal is to reconstruct all of the ancestral genomes so that the number
of evolutionary operations is minimized.
Algorithms for reconstructing evolutionary history from gene orders are usually based
on repeatedly computing medians of genomes at neighbouring vertices of the tree. We propose a new,
more general approach, based on an iterative local optimization procedure.
In each step, we propose candidates for ancestral genomes and choose the best ones by dynamic programming.
We have implemented our method and used it to reconstruct evolutionary history
of 16 yeast mtDNAs and 13~\emph{Campanulaceae} cpDNAs.

\end{abstract}

\section{Introduction}

Phylogeny is an evolutionary history of a group of organisms, and
is usually represented by a phylogenetic tree, where leaves represent
extant species and internal nodes represent their ancestors.


During the evolution, the genomes of species undergo various mutations. Large-scale mutations, such as
inversions or transpositions, change positions of genes. Other evolutionary events may change
the number of chromosomes or their topology (linear to circular and vice versa). Since these large-scale
mutations are much rarer than the point mutations, they constitute a valuable source of information
for phylogenetic analysis.

We can define a distance between two genomes as the minimum number of evolutionary operations needed
to transform one genome into the other. Then a parsimony approach to phylogeny reconstruction 
is to minimize the total number of evolutionary operations throughout the evolution.
More precisely: Given the gene orders of extant species, the goal is to find a phylogenetic tree
together with gene orders of the ancestral species, that minimizes the number of evolutionary operations
-- this is the \emph{large phylogeny} problem.

In this paper, we are interested in the \emph{small phylogeny} problem, where the phylogenetic tree
is given, and the goal is only to reconstruct the ancestral genomes.
This problem is interesting in its own right. Furthermore, the small phylogeny problem is a subproblem
that needs to be solved when solving large phylogeny problem. The existing programs reconstruct
the phylogenetic tree either by listing all tree topologies and solving the small phylogeny problem
(as in \prog{GRAPPA} software \cite{MoretW02}) or by incrementally (heuristically) adding new branches and 
solving the small phylogeny problem (as in \prog{MGR} software \cite{BourqueP02}).


In Section 2, we define our problem and summarize the previous work on the problem.
Section 3 describes our new method for reconstructing evolutionary histories.
In Section 4, we present results on real datasets -- yeast mitochondrial and plant chloroplast genomes,
and we draw conclusions in Section 5.
\section{Preliminaries}

\subsection{Genome Models}

Various genome models have been proposed:
\begin{itemize}
\item Do we know on which strand the markers (genes, synteny blocks) are located?
      (Unsigned vs.\ signed models)
\item Do the species in consideration have a single chromosome or multiple chromosomes?
      (Unichromosomal vs.\ multichromosomal models)
\item Do the species have linear chromosomes, circular chromosomes, or a mix of both?
      (Linear vs.\ circular vs.\ mixed models)
\item What are the operations that rearrange the genomes throughout the evolution?
      Reversals? Transpositions? Translocations?
      Fusions and fissions? Some combination of the former? 
\end{itemize}

\begin{definition}[Genome model]
Genome model is a pair $(\G,d)$, where $\G$ is the set of all possible genomes and
$d$ is a distance measure on $\G$.
\end{definition}

\ex 1 One simple option is to model (unsigned unichromosomal) genome as a permutation $\pi$ of markers $\{1,2,\ldots,n\}$.
Let $\pi,\rho$ be two genomes (permutations). For convenience, let us define $\pi_0=\rho_0=0$ and $\pi_{n+1}=\rho_{n+1}=n+1$.
Now, if $\pi_i$ and $\pi_{i+1}$ are two consecutive markers in $\pi$,
which are \emph{not} consecutive in $\rho$, we call the pair $(\pi_i,\pi_{i+1})$ a \emph{breakpoint}. The breakpoint
distance between $\pi$ and $\rho$ is simply the number of breakpoints in $\pi$.

\ex 2 If we know the orientation of markers, we can model genomes as signed permutations,
where each marker $g$ has orientation $\lo g$ or $\ro g$. By $-g$ we denote marker with the opposite
orientation (i.e., $-\lo g=\ro g$ and $-\ro g=\lo g$). Let $\pi=(\pi_1,\pi_2,\ldots,\pi_n)$ be a signed
permutation. \emph{Reversal} operation $\hbox{rev}(i,j)$ transforms $\pi$ into 
$$\pi'=(\pi_1,\ldots,\pi_{i-1},-\pi_j,-\pi_{j-1},\ldots,-\pi_i,\pi_{j+1},\ldots,\pi_n).$$

In the \emph{reversal model}, $\G$ is the set of all signed permutations and the distance $d$ between genomes $\pi,\rho$
with the same marker content is the minimum number of reversals needed to transform $\pi$ into $\rho$.
Reversal distance can be computed in linear time \cite{BaderMY01}.

\ex 3 In the \emph{double-cut and join (DCJ) model} \cite{YancopoulosAF05,BergeronMS06},
we represent a genome as a graph consisting of cycles and paths (representing circular and linear chromosomes,
respectively). Each (oriented) marker is represented by two
vertices, called \emph{extremities} of the marker; the ends of linear chromosomes are represented
by special vertices called \emph{telomeres}. The edge set of this graph consists of the \emph{marker
edges}, joining the two extremities of each marker, and the \emph{adjacencies}, joining two
consecutive extremities in the genome or an extrimity with a telomere.


A DCJ operation takes two adjacencies, $\{p,q\}$ and $\{r,s\}$, and replaces them by either $\{p,r\}$ and $\{q,s\}$,
or $\{p,s\}$ and $\{q,r\}$. This operation is quite general. A single DCJ operation can represent a reversal, translocation,
fusion, fision, excision, or integration of a circular chromosome. Two operations can simulate a transposition.
The DCJ distance is defined as the minimum number of DCJ operations needed to
transform one genome into another. The distance can be computed in linear time \cite{BergeronMS06}.

\ex 4 In the \emph{Hannenhalli-Pevzner (HP) model} \cite{HanennhalliP95}, only genomes with linear chromosomes
are allowed. The evolutionary operations are reversals, translocations, fusions and fissions. This model can be seen
as a DCJ model restricted to linear chromosomes (with operations that do not create circular chromosomes).
HP-distance can also be computed in linear time \cite{BergeronMS08,BergeronMS09}.

\subsection{Rearrangement Phylogenies} 

In the small phylogeny problem, we are given genomes of the extant species together
with a phylogenetic tree, and the goal is to compute genomes of their ancestors.
According to the parsimony principle, the best reconstruction is such that
involves the smallest number of rearrangement operations in the evolution.

More specifically, we choose some representation of genomes and genomic distance,
and we try to find such ancestral genomes that the sum of the distances along the edges
of the phylogenetic tree is minimized.

More formally, let $T=(V,E)$ be a phylogenetic tree with the set of leaves~$L$.
For each leaf, we are given a genome of the corresponding species, i.e., we are given
a function $g:L\to\G$. An \emph{evolutionary history} is a function $h:V\to \G$
extending $g$ that maps a genome to each vertex.

\begin{probl}[Small phylogeny problem]
Given a phylogenetic tree $T=(V,E)$ and genomes of the extant species, find an
evolutionary history $h$ with the minimum score
$$ d(h) = \sum_{\{u,v\}\in E} d(h(u),h(v)). $$
\end{probl}

A special case of a phylogeny problem with only three extant species is called \emph{median problem}.
For three species, there is only one unrooted phylogenetic tree -- a star with three edges. The only
ancestor is called \emph{median}.

\begin{probl}[Median problem]
Given three genomes $\Pi_1$, $\Pi_2$, and $\Pi_3$, find the median genome $\Pi_M$,\
such that the sum of distances from $\Pi_M$ to each genome 
$$ d(\Pi_1,\Pi_M) + d(\Pi_2,\Pi_M) + d(\Pi_3,\Pi_M) $$
is minimized.
\end{probl}

Median problem has been shown to be NP-hard for almost every considered genome model
(unichromosomal reversal distance \cite{Caprara03}, unichromosomal breakpoint
distance \cite{PeerS98,Bryant98}, multichromosomal linear breakpoint distance
\cite{TannierZS09}, unichromosomal \cite{Caprara03} and multichromosomal \cite{TannierZS09}
DCJ distance) and is conjectured to be NP-hard for other genome models.
One notable exception is the breakpoint distance on multiple circular or mixed chromosomes
for which a median can be computed in polynomial time \cite{TannierZS09}.

Note that NP-hardness of the median problem also implies NP-hardness of the small phylogeny
problem. (The complexity of the small phylogeny problem under the breakpoint distance 
is unknown.)

\subsection{Previous Work}\label{ss:prev}

Despite the fact that median problem has been proven to be NP-hard, the prevailing approach to solving
small phylogeny problem is based on solving the median problem exactly or heuristically.
The so called \emph{steinerization} method \cite{SankoffCL76} iteratively improves the evolutionary history
until a local optimum is reached. In each iteration, we go through all internal vertices $v$.
We take an ancestral genome $\Pi_v$ and its three neighbours $\Pi_a,\Pi_b,\Pi_c$, we compute
a median $\Pi_M$ of these neighbours, and if it has a better score than $\Pi_v$, we replace $\Pi_v$ by $\Pi_M$.
If no vertex can be improved by taking the median of its neighbours, we have a locally optimal
evolutionary history.

This approach was initiated by Blanchette, Bourque, and Sankoff \cite{BlanchetteBS97,SankoffB97}
and the method was implemented in the \prog{BPAnalysis} software. \prog{BPAnalysis} solves
the large phylogeny problem under the breakpoint distance by generating all
tree topologies and solving the small phylogeny problem. Blanchette, Bourque,
and Sankoff \cite{BlanchetteBS97,SankoffB97} reduce breakpoint median problem to travelling salesman
problem and then use a branch-and-bound algorithm to solve the resulting instance
of TSP exactly.

\prog{GRAPPA} software by Moret et al.\ \cite{MoretWBWY01,MoretWWW01,MoretTWW02} solves both breakpoint phylogeny
and reversal phylogeny problem. Breakpoint median problem is solved using an approximate
TSP solver. For the reversal median problem solvers, Siepel and Moret \cite{SiepelM01} and
Caprara \cite{Caprara01} are used. The package also contains an exact exponential algorithm
by Tang and Moret \cite{TangM05} and a heuristic solver by Arndt and Tang \cite{ArndtT08}.
An extension for transposition phylogeny problem was proposed by Yue et al.\ \cite{YueZT08}

Another approach to the large phylogeny problem by Bourque and Pevzner \cite{BourqueP02}
is the \emph{sequential addition} heuristic. Instead of generating all topologies,
the method tries to build one or several trees with a small score. In this heuristic,
we build the tree incrementally by adding new species. In each step, we choose an edge
to be split and replaced by a 3-star with the new species. We choose the best edge
greedily so that the resulting intermediate tree has minimum score.

This method is implemented in the \prog{MGR} software. The reversal median problem
is solved heuristically by moving genomes closer to each other. The program was also
reimplemented by Adam and Sankoff \cite{abc} using the DCJ model and heuristic DCJ
median solver.

More recently, Xu and Sankoff \cite{XuS08} proposed a fast exact DCJ median solver.

\section{Iterative Local Optimization}

Here, we propose a new general approach based on iterative local optimization.
The basic idea is that in each step, we propose candidates for ancestral
genomes and choose the best combination of the candidates by dynamic programming.

In particular, consider a phylogenetic tree $T=(V,E)$ with the set of leaves $L$
and genomes of extant species $g:L\to\G$. Let $\H$ be the set of all possible 
evolutionary histories. We start with some history $h_0$.
For a particular history $h$ and each internal vertex $v$, we propose a set of candidates $\cand(h,v)$.
We define a neighbourhood of history $h$ as $N(h) = \{ h' \mid \forall v\in V: h'(v)\in \cand(h,v)\}$,
i.e., we consider all the possible combinations of candidate genomes as neighbouring histories.
We then find the best history in the neighbourhood by a dynamic programming algorithm and if the
new history is better than the previous one, we take it and repeat the iteration.
Otherwise, we have found a local minimum and the algorithm terminates.

$$\begin{algorithm}[H]
\KwData{evolutionary history $h$}
\KwResult{local optimum}
$s' \gets score(h)$, $s \gets \infty$ \;
\While{$s'<s$}{
  $\cand \gets \null$generate lists of candidates (neighbourhood of $h$)\;
  $h \gets best(\cand)$\;
  $s \gets s'$, $s'\gets score(h)$\;
}
\Return $h$
\caption{Local optimization}
\end{algorithm}$$

\ex 1 For each internal vertex $v$, the set of candidates can be all the genomes within the distance $1$
from $h(v)$. The neighbourhood of $h$ is the set of all histories we can obtain from $h$ by performing
at most one operation to each ancestral genome. (Note that the size of $N(h)$ is exponential in the number
of internal vertices, but as we will see later, we will never require to enumerate the whole neighbourhood.)

\ex 2 The steinerization approach mentioned in Section~\ref{ss:prev} is a special case of our method:
Here, $\cand(h,v)=\{h(v)\}$ for all vertices except for one vertex $w$ with neighbours $a,b,c$, for
which $\cand(h,w)=\{h(w),\mathop{\hbox{median}}(h(a),\break h(b),h(c))\}$.

\subsection{Finding the Best History in a Neighbourhood}

Even though the size of the neighbourhood can be exponential (it has $\prod_v |\cand(v)|$ elements),
the best history can be found in polynomial time using dynamic programming.

Let $c^{(u)}_i$ be the $i$-th candidate from $\cand(h,u)$ and let $M[u,i]$ be the lowest
score we can achieve for the subtree rooted at $u$ if we choose the candidate $c^{(u)}_i$ as an ancestor.
$M[u,i]=0$ if $u$ is a leaf. If $u$ is an internal vertex with children $v$ and $w$, we first compute
values $M[v,j]$, $M[w,k]$ for all $j,k$. Then
$$ M[u,i] = \min_j \{ M[v,j]+d(c^{(u)}_i,c^{(v)}_j) \} + \min_k \{ M[w,k]+d(c^{(u)}_i,c^{(w)}_k) \}. $$
This could be easily generalized for phylogenetic trees that are not binary. 

If $n$ is the number of species, $m$ is the number of markers in genomes, and $k$ is the number of candidates
for each ancestor, the best history can be found in time $O(nmk^2)$ (provided that the distance between
two genomes can be computed in $O(m)$ time). 


\subsection{Strategies for Proposing Candidates}

There are many methods for proposing candidates. In general, by proposing more candidates we
explore larger neighbourhood, but finding the best choice of candidates is slower. Furthermore,
if we propose only candidates that are close to the genomes in the current history, the convergence
to the local optimum may be slow.
Here, we list several strategies for proposing candidates.

{\it Descendants.} In the initialization step, we can take genomes of the extant species as candidates
to get an evolutionary history to begin with.

{\it Neighbours.} We already mentioned that we can include neighbourhoods of individual genomes,
i.e. $N(\Pi) = \{ \Gamma \mid d(\Pi,\Gamma)\leq 1 \}$, in the set of candidates.
For most models the size of $N(\Pi)$ is roughly quadratic in the number of markers.

{\it Scenarios.} For vertex $v$ with neighbouring vertices $u$ and $w$, we can take intermediate
genomes as candidates, i.e.\ if $\Pi,\Gamma$ are genomes at $u$ and $w$, we can sample genomes
$\Theta$ such that $d(\Pi,\Theta)+d(\Theta,\Gamma)=d(\Pi,\Gamma)$.

{\it Medians.} In the steinerization method, we have a median of genomes in the neighbouring vertices
as a candidate. Note that often there are many medians with the same score. In our method, we do not
need to decide which median to use, but instead we can consider all medians as candidates.

If we compute median by branch-and-bound technique, the time to list all medians is comparable
to the time to find just one.
If we try to find median heuristically by moving the given genomes closer and closer to each other,
we can take the intermediate genomes as candidates.
Another option is to find just a single median and then search its neighbourhood.


{\it Layers.} Our method runs in time quadratic in the size of candidate sets. That means, that the
algorithm is slow for large sets of candidates.
However, the $O(nmk^2)$ bound is tight only when all the candidate sets contain $k$ genomes.
Another variation of our method is to divide the vertices into layers by their depth.
We can generate large sets of candidates for the odd layers and small sets of candidates for even layers
(or vice versa).
If the corresponding sizes are $k_1$ and $k_2$, the running time will be $O(nmk_1k_2)$.

{\it Best histories.} We can take locally optimal histories and put the reconstructed ancestors into
the sets of candidates. This way we can "crossbreed" locally optimal solutions discovered previously.

\subsection{Dealing with Unequal Gene Content}\label{s:uneqgc}

Genome models usually do not account for duplications or losses since the distance is
then difficult to compute. In our method, it is easy to deal with a few duplications or losses
by considering different forms of genomes of the extant species.
Instead of running the algorithm many times for different choices of extant genomes,
we can just put the alternative genomes in sets of candidates.
One of the alternative gene orders is then chosen in each leaf as a representative,
so as to minimize the overall parsimony cost.

For example, if a few of the genomes contain only a few duplications, we can consider
all forms where we delete all but one of the copies as candidates in the corresponding leaves.
On the other hand, if all of the genomes contain a few duplications, we can consider all
the differentiated genomes, where we treat each copy as a different marker.


\section{Experiments on Real Data} 

\subsection{The \emph{Hemiascomycetes} mtDNA Dataset}

\begin{figure}[ht]
\centering
\includegraphics[width=0.55\textwidth]{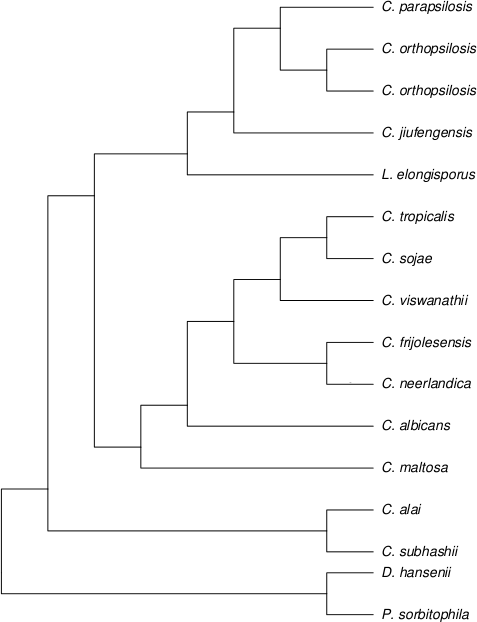}
\caption{Phylogenetic tree of CTG clade of \emph{hemiascomycetes}}
\label{fig:tree1}
\end{figure}

We have implemented our method using the most general DCJ model
and used it to study 
evolution of gene order in 16 mitochondrial
genomes from the ’CTG’ clade of \emph{hemiascomycetes}. The phylogenetic
tree was calculated by \prog{PhyloBayes} (Fig.~\ref{fig:tree1}) and is
supported by high posterior probabilities on most branches.
The tree is also consistent with the study of Fitzpatrick et al.\ \cite{FitzpatrickLSB06}.

The genomes in the dataset consisted of 25 markers (synteny blocks) comprising
14 protein-coding genes, two ribosomal RNA genes and around 24 tRNAs.
Genomes of \emph{C. subhashii, C. parapsilosis}, and \emph{C. orthopsilosis} are linear;
\emph{C. frijolesensis} has two linear chromosomes; other considered species have circular-mapping
chromosomes -- thus a general model such as DCJ was needed.

The DCJ model does not handle genomes with duplicated genes. To resolve
recent duplications in some of the genomes (\emph{C. albicans, C. maltosa,
C. sojae, C. viswanathii}), we removed the duplicated genes, and included
both possible forms of the genomes as alternatives in the corresponding
leaves as described in Section~\ref{s:uneqgc}. Similarly, both isomers
are allowed in the genomes that include long inverted repeats
(\emph{C. alai, C. albicans, C. maltosa, C. neerlandica, C. sojae, L. elongisporus}).

We penalized occurrences of multiple circular chromosomes, or combinations of
linear and circular chromosomes in ancestral genomes.

\begin{figure}[p]
\centering
\includegraphics[angle=90,width=0.7\textwidth]{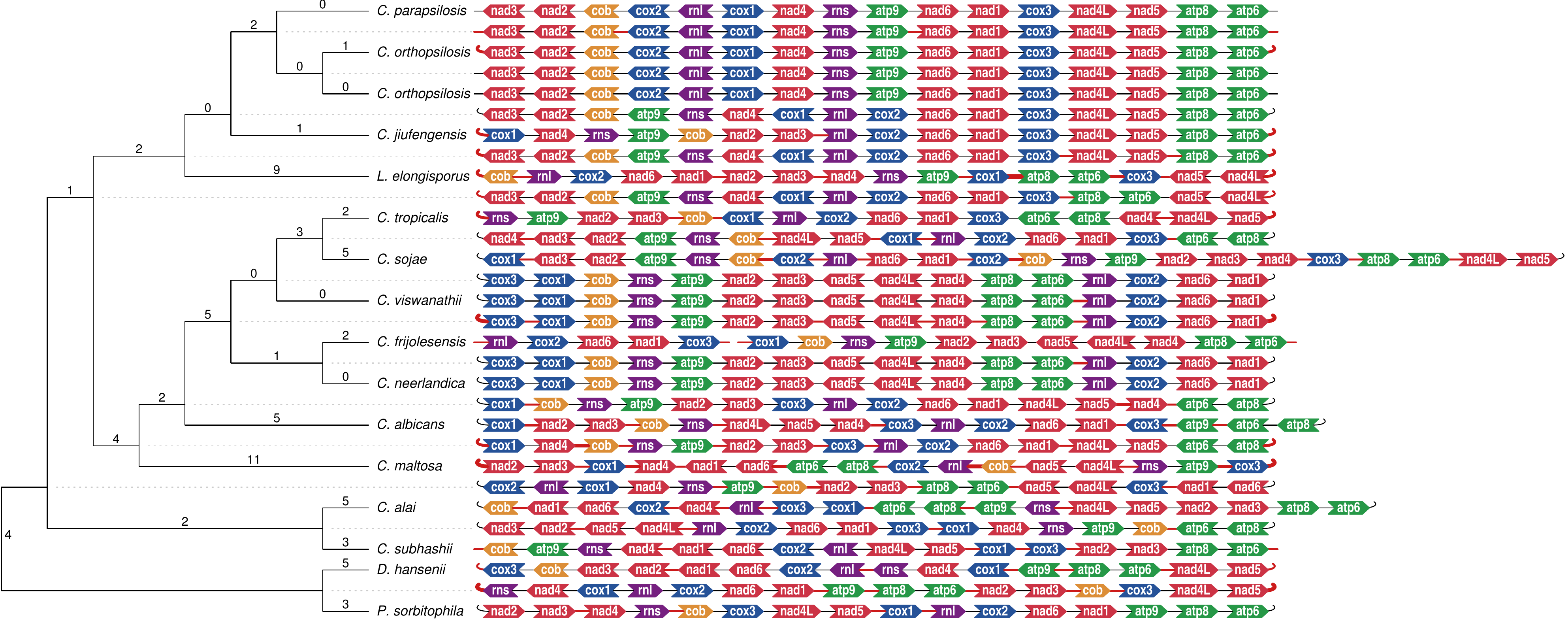}
\caption{Reconstructed evolutionary history of hemiascomycetes' mitochondrial DNA}
\label{fig:hist1}
\end{figure}

We found evolutionary history with 78 DCJ operations (Fig.~\ref{fig:hist1})
which agrees with manually reconstructed events among closely related species \cite{Valach10}.
This study shows practical applicability of our approach to real biological datasets.

\subsection{The \emph{Campanulaceae} cpDNA Dataset}

To compare our method with existing ones, we also applied our program
to a well-studied dataset of \emph{Campanulaceae} chloroplast genomes \cite{Cosner00}.
This dataset consists of 13 species, each genome consists of one
circular chromosome with 105 markers.

\begin{figure}[ht]
\centering
\includegraphics[width=0.65\textwidth]{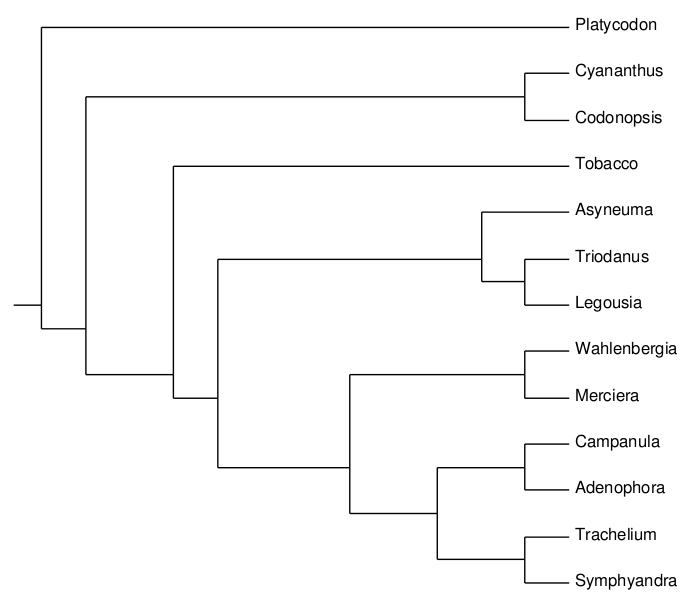}
\caption{Phylogenetic tree of \emph{Campanulaceae}}
\label{fig:tree2}
\end{figure}

Using GRAPPA software, Moret et al. \cite{MoretWWW01} found 216 tree topologies
and evolutionary histories with 67 reversals. Bourque and Pevzner \cite{BourqueP02}
using MGR later found a solution with 65 reversals. Their phylogenetic tree is shown
in Fig.~\ref{fig:tree2}.

Adam and Sankoff \cite{abc} used the more general DCJ model and the phylogenetic tree
by Bourque and Pevzner. They found an evolutionary history with 64 DCJ operations
if the ancestral species were required to have a single chromosome and a history
with 59 DCJ operations if the ancestral species were unconstrained.
However, as Adam and Sankoff note: "There is no biological evidence in the \emph{Campanulaceae},
or other higher plants, of chloroplast genomes consisting of two or more circles."
The additional circular chromosomes are an artifact of the DCJ model, where a transposition
or a block interchange operation can be simulated by circular excision and reincorporation.

We ran our program on this data set penalizing multiple chromosomes and we have found
several evolutionary histories with 59 DCJ operations, where all the ancestral species
had single circular chromosomes.

\section{Conclusion}

In this paper, we have developed a new method for reconstructing evolutionary history
and ancestral gene orders, given the gene orders of the extant species and their phylogenetic tree.
We have implemented our method using the double-cut and join model and used it to study
evolution of gene order in 16 mitochondrial yeast genomes. The study shows practical applicability
of our approach to real biological datasets. We have also analyzed the well-studied \emph{Campanulaceae}
dataset and we have improved upon the results of Adam and Sankoff \cite{abc}.

\subsubsection*{Acknowledgements.}
This research was supported by Marie Curie reintegration grants
IRG-224885 and IRG-231025 to Tom\'a\v s Vina\v r and Bro\v na Brejov\'a,
and VEGA grant 1/0210/10.

\bibliographystyle{splncs}
\bibliography{main}

\end{document}